\documentstyle[eqsecnum,aps]{revtex}

\begin{document}
\title{A Study of the Coulomb Dissociation of $^8B$ and the $^7Be(p, \gamma )^8B$
Reaction}
\author{Carlos A. Bertulani$^a$, and Moshe Gai$^b$}
\address{
$^a$ Instituto de F\'\i sica, Universidade Federal
do Rio de Janeiro\\
21945-970 \ Rio de Janeiro, RJ, Brazil. E-mail:
bertu@if.ufrj.br\\
$^b$
Department of Physics, U46, University of Connecticut\\
2152 Hillside Rd., Storrs, CT 06269-3046, USA. E-mail:
gai@uconnvm.uconn.edu}
\date{\today }
\maketitle
\begin{abstract}
We study the breakup reactions of $^8B$ projectiles in high energy (50 and
250 MeV/u) collisions with heavy nuclear targets ($^{208}Pb$). The intrinsic
nuclear wave functions are calculated using a simple model, as well as a
simple optical potential. We demonstrate that nuclear effects are negligible
and evaluate the contributions of various (E1, E2 and M1) multipolarities. A
good agreement with measured data is obtained with insignificant M1
contribution (at 50 MeV/u) and very small E2 contribution.
\end{abstract}

\bigskip



\section{Introduction}

The use of the Coulomb dissociation method \cite{Ba86,Be88} has proven to be
a useful tool for extracting radiative capture reaction cross section of
relevance for nuclear astrophysics. In particular it appears that the
Coulomb dissociation of $^8B$ is very useful \cite{Ga95} for elucidating the
most uncertain nuclear input to the standard solar model - the formation of $%
^8B$ via the $^7Be(p,\gamma)^8B$ reaction. However, a few lingering
questions still need to be addressed, including the importance of nuclear
excitations for the kinematics of the RIKEN \cite{Mo94} and GSI \cite{Sue96}
experiments, performed at approximately 50 and 250 MeV/u, respectively, as
well as the relative importance of the various E1, E2 and M1 electromagnetic
excitations. In this paper we attempt to resolve these issues by using a
relatively simple but still realistic nuclear model that however yields a
very good agreement with data and suggest that nuclear excitations as well
as E2 and M1 excitations are negligible for the kinematical conditions of
the RIKEN and GSI experiments.

\section{Transition densities}

We use for $^8B$ a similar model as in ref. \cite{Ber96}, assuming that the $%
J_0=2^+$ ground state can be described as a $j_0=p_{3/2}$ proton coupled to
the $I_c=3/2^-$ ground state of the $^7Be$ core. The spectroscopic factor
for this configuration was taken as unity. The single particle states, $%
\Psi_{JM}$, are found by solving the Schr\"odinger equation with spin-orbit
term and matching to asymptotic Coulomb waves. The parameters of the
potentials are given in table 1 of ref. \cite{Ber96}.

A multipole expansion of the transition density, $\delta \rho$, yields 
\begin{equation}
\delta \rho ({\bf r})\equiv \Psi_{JM}^*({\bf r})\; \Psi_{J_0M_0} ({\bf r})=
\sum_{\lambda\mu} \delta \rho_{\lambda\mu}^{(JM)} (r) \; Y_{\lambda \mu} (%
\hat{{\bf r}}) \ ,
\end{equation}
where 
\begin{equation}
\delta \rho_{\lambda\mu}^{(JM)}(r) = \int \Psi_{JM}^* ({\bf r})\;
Y_{\lambda\mu}^*(\hat{{\bf r}}) \; \Psi_{J_0M_0}({\bf r})\; d\Omega \ ,
\end{equation}
and $|J_0 M_0>\ (|JM>)$ denote the initial (final) state of the projectile.

The electromagnetic transition matrix for the multipolarity $\pi \lambda \mu 
$ is given by 
\begin{eqnarray}
\left\langle JM\left| {\cal M}_{\pi \lambda ,-\mu }\right|
J_0M_0\right\rangle &=&\int dr\;r^\lambda \;\delta \rho _{E\lambda \mu
}^{(JM)}\ \ \ \ \ {\rm (electric)};  \nonumber \\
&=&\int dr\;r^{\lambda -1}\;\delta \rho _{M\lambda \mu }^{(JM)}\ \ \ \
\ {\rm (magnetic)}\ ,
\end{eqnarray}
where, for electric multipole transitions 
\begin{eqnarray}
\delta \rho _{E\lambda \mu }^{(JM)}(r) &=&{\frac 1{2\sqrt{4\pi }}}%
\;(-1)^{I_c+J_0+\lambda +j_0+l+l_0+\mu }\;e_{_E}\;{\frac{\hat{J_0}\hat{%
\lambda}\hat{j_0}}{\hat{j}}}\;  \nonumber \\
&\times &\left\langle J_0M_0\lambda \mu \Big| JM\right\rangle \left\langle
j_0{\frac 12}\lambda 0\Big| j{\frac 12}\right\rangle \;\Biggl\{ {%
{j \atop J_0}
}{%
{J \atop j_0}
}{%
{I_c \atop \lambda }
}\Biggr\} \;\delta \rho _{E_xlj;l_0j_0}^{(J)}\ ,
\end{eqnarray}
where $\hat{l}=\sqrt{2l+1}$, $e_E=e\Big[ (1/2)^\lambda +(-1/8)^\lambda \Big]$%
, and 
\begin{equation}
\delta \rho
_{E_xlj;l_0j_0}^{(J)}(r)=r^2\;R_{E_xlj}^{(J)*}(r)\;R_{l_0j_0}^{(J_0)}(r)\ .
\end{equation}
Above, $R_{l_0j_0}^{(J_0)}(r)$ is the radial wave function for the
ground-state and $R_{E_xlj}^{(J)}(r)$ is the radial wave function for a
state in the continuum with excitation energy $E_x$.

For magnetic dipole transitions 
\begin{eqnarray}
\delta_{M\lambda\mu}^{(JM)} (r)& =& \mu_{_N} \ (-1)^{j_0+I_c+J_0+1} \ {\frac{%
\hat{J_0} }{\hat{l_0}}} \ \left\langle J_0M_0\lambda\mu | JM\right\rangle\ %
\Biggl\{ {%
{j  \atop J_0}
}{%
{J  \atop j_0}
} {%
{I_c  \atop 1}
} \Biggr\}  \nonumber \\
&\times& \Biggl\{e_{_M}\ \bigg[ 2 {{\frac{\tilde{j_0}}{\hat{l_0}}}\ \Big( %
l_0 \delta_{j_0,l+{\frac{1}{2}}} +(l+1)\delta_{j_0,l-{\frac{1}{2}}}\Big)%
+(-1)^{l_0+{\frac{1}{2}}-j} \ {\frac{\hat{j_0}}{\hat{2}}}\delta_{j_0,l_0\pm {%
\frac{1}{2}}}\delta_{j,l\mp {\frac{1}{2}}}} \bigg]  \nonumber \\
&+& g_p \ \bigg[ (-1)^{l_0+{\frac{1}{2}} - j_0}\ {\frac{\tilde{j_0}}{\hat{l_0%
}}} \delta_{j_0,j}-(-1)^{l_0+{\frac{1}{2}} - j}\ {{\frac{\hat{j_0}}{\hat{2}%
\hat{l_0}}}\ \delta_{j_0,l\pm {\frac{1}{2}}}\delta_{j,l\mp {\frac{1}{2}}}%
\bigg]\Biggr\} \delta \rho_{E_xlj;l_0j_0}^{(J)}(r)}  \nonumber \\
&+&\mu_c\ (-1)^{I_c+j_0+J+1} \ \left\langle J_0M_0\lambda\mu |
JM\right\rangle \ \hat{J_0}\hat{J}\hat{I_c}\tilde{I_c} \ \biggl\{ {%
{I_c  \atop J_0}
}{%
{J  \atop I_c}
} {%
{j_0  \atop 1}
} \biggr\}\ \delta \rho_{E_xlj;l_0j_0}^{(J)}(r)  \nonumber \\
\end{eqnarray}
where $e_M=3/2$, $\mu_c=-1.7 \ \mu_N$, $\mu_N=e\hbar/2m_pc$, $m_p$ is the
proton mass, and $\tilde{l}=\sqrt{l(l+1)}$.

In addition to the electric and magnetic transitions due to Coulomb
excitation of the projectile, we will also consider the transitions induced
by the nuclear field of the target in peripheral collisions. The nuclear
induced transition density is built also as in eqs. (1,2), and for isoscalar
excitations (we will not consider nuclear isovector excitations for reasons
explained below) we can write the transition density as $\delta
\rho_{N\lambda\mu}^{(JM)}= \delta \rho_{E\lambda\mu}^{(JM)} /e_{_E}$.

The transition densities, $\delta \rho _{E_xlj;l_0j_0}^{(J)}(r)$, are shown
in figures 1(a) and 1(b). In figure 1(a) we show the transition densities to
the $J=1^{+}$, and $J=3^{+}$, resonances at $E_x=0.63$ MeV, and $E_x=2.17$
MeV, respectively. For convenience, the transition densities have been
integrated over the width of the resonance. Also shown in this figure
(dotted line) is the transition density obtained by the Tassie model (see,
e.g., ref. \cite{Sat83}): $\delta \rho _{Tassie}(r)=(\beta r)\;d\rho /dr$,
where $\rho (r)$ is the ground-state density of the $^8B$, taken as $\rho
(r)=\rho _0\left[ 1+\alpha (r/a)^2\right] \;\exp \{-r^2/a^2\}$, with $\alpha
=0.631$ fm and $a=1.77$ fm. The deformation parameter, $\beta $, is chosen
so that the Tassie transition density is normalized to the peak of the
transition density to the $J=3^{+}$ state. We see that the transition
densities extend further out than the Tassie transition density. This is
even more visible for the transition densities to the non-resonant p- and
f-waves, as shown in figure 1(b), for $E_x=350\ keV$. These transition
densities extend to very large radial values. This is an important result
since it invalidates calculations based on the Tassie model for the
transition densities in the case of halo nuclei, as was pointed out in ref. 
\cite{BS95}.

\section{Optical and transition potentials}

Since there is no data for the elastic scattering of $^8B$ on $Pb$ targets
at the energies that we want to consider, we construct an optical potential
using an effective interaction of the M3Y type \cite{KBS84,BLS97} modified
so as to reproduce the energy dependence of total reaction cross sections,
i.e. \cite{BLS97}, 
\begin{equation}
t(E,s)=-i{\frac{\hbar v}{2 t_0}} \; \sigma_{NN}(E) \;
\left[1-i\alpha(E)\right] \; t(s) \ ,  \label{tes}
\end{equation}
where $t_0=421$ MeV is the volume integral of the M3Y interaction $t(s)$, $v$
is the projectile velocity, $\sigma_{NN}$ is the nucleon-nucleon cross
section, and $\alpha$ is the real-to-imaginary ratio of the forward
nucleon-nucleon scattering amplitude. The optical potential is given by 
\begin{equation}
U(E,{\bf R})=\int d^3r_1\; d^3r_2 \; \rho_{_P}({\bf r}_1) \rho_{_T}({\bf r}%
_2) \; t(E,s) \ ,
\end{equation}
where ${\bf s}={\bf R}+{\bf r}_2-{\bf r}_1$, and $\rho_{_T}$ ($\rho_{_P}$)
is the ground state density of the target (projectile).

According to this model, the optical potential is about two times smaller at
250 MeV/nucleon compared to 50 MeV/nucleon. The optical potentials generated
in this way will be used to obtain the distorted waves in the eikonal
approximation (see next section).

The transition potential for excitations of the projectile is given by 
\begin{equation}
\Delta U(E,{\bf R})=\int d^3r_1\; d^3r_2 \; \delta \rho_{_P}({\bf r}_1)
\rho_{_T}({\bf r}_2) \; t(E,s) \ ,
\end{equation}
A multipole expansion of the transition potential yields 
\begin{equation}
\Delta U(E, {\bf R})=\sum_{\lambda\mu} \delta U_{\lambda\mu}^{(JM)} (E, R)
\; Y_{\lambda\mu} (\hat{{\bf R}}) \ .
\end{equation}
where 
\begin{equation}
\delta U_{\lambda\mu}^{(JM)}(E, R) =\int dr \; r^2 \; I_\lambda(E, R, r) \;
\delta\rho_{N\lambda\mu}^{(JM)}(r) \ ,
\end{equation}
with 
\begin{equation}
I_\lambda(E,R, r_2)=(2\pi)^2 \int d (\cos\theta_1) \; d (\cos\theta_2)\;
dr_1 \; r_1^2 \; t(E,s)\; P_\lambda(\cos\theta_2) \ ,
\end{equation}
where $\theta_1$ is the angle between ${\bf r}_1$ and ${\bf R}+{\bf r}_2$,
and $\theta_2$ is the angle between ${\bf r}_2$ and ${\bf R}$.

In figures 2(a) and 2(b) we show the transition potentials 
\[
\delta U_{E_x l j;l_0j_0} (R)=\int dr \; r^2 \; I_2 (R,\;r)\;
\delta\rho_{E_xlj;l_0j_0}(r) 
\]
as a function of the radial distance, for $^8B$ projectiles incident on $Pb$
at E=50 MeV/nucleon. In figure 2(a) we show the real part of the transition
potentials, integrated over the width of the $J=1^+$ (solid line) and $J=3^+$
(dashed line) resonances. They are compared with the Copenhagen model
(dashed line) for the transition potential (called by ``standard potential"
in ref. \cite{Sat83}), i.e., $\delta U_{Cop.}= (\beta r)\; dU_{opt}/dr$,
with an arbitrary value for the deformation parameter $\beta$. Again we see
that the transition potentials have a quite different radial dependence than
the Copenhagen transition potential model. For the non-resonant p and f
waves this difference is even more pronounced, as we see in figure 2(b), for 
$E_x=350 \ keV$. According to eqs. (3.1, 3.3), at 250 MeV/nucleon the
transition potentials have the same shape as those presented in figures 2(a)
and 2(b), but are about twice smaller in magnitude since the effective
interaction $t(E,s)$ is reduced by nearly the same amount.

\section{Nuclear and Coulomb excitation cross sections}

The amplitude for the nuclear excitation of high energy projectiles is given
by 
\begin{eqnarray}
f_N&=&-{\frac{\mu_{_{PT}} }{2 \pi \hbar^2}} \; \int d^3 R \; \Psi^{(-)*}(%
{\bf R}) \; \Psi^{(+)} ({\bf R}) \; \Delta U({\bf R})  \nonumber \\
&=&\sum_{\lambda\mu} f_{N\lambda\mu}^{(JM)} \ ,
\end{eqnarray}
where $\mu_{_{PT}}$ is the reduced mass of the projectile+target system, $%
\Psi^{(-)}$ ($\Psi^{(+)}$) is the incoming (outgoing) scattering wave of the
system, and 
\begin{equation}
f_{N\lambda\mu}^{(JM)}=-{\frac{\mu_{_{PT}} }{2 \pi \hbar^2}} \; \int d^3 R
\; \Psi^{(-)*}({\bf R}) \; \Psi^{(+)} ({\bf R}) \; \delta
U_{\lambda\mu}^{(JM)}({\bf R})\; Y_{\lambda\mu}(\hat{{\bf R}}) \ .
\end{equation}

For high energy projectiles, we can use the eikonal approximation ($Q=2k\sin
(\theta/2$)), 
\begin{equation}
\Psi^{(-)*}\Psi^{(+)} = \exp\left\{i{\bf Q.R}+i\chi(b)\right\} \ ,
\end{equation}
with the eikonal phase given by 
\begin{equation}
\chi(b)=2\eta \ln (kb)-{\frac{1}{\hbar v}} \; \int_{-\infty}^\infty dz \;
U_{opt}(R) \ ,
\end{equation}
where $\eta=Z_PZ_Te^2/\hbar v$, $k$ is the projectile momentum, and $R=\sqrt{%
b^2+z^2}$. The optical potential, $U_{opt}$, in the above equation is given
by eq. (3.2).

Following ref. \cite{BN93}, the Coulomb amplitude is given by 
\begin{equation}
f_C=\sum_{\lambda \mu }f_{C\lambda \mu }\ ,
\end{equation}
where 
\begin{eqnarray}
f_{C\lambda \mu } &=&i^{1+\mu }\;{\frac{Z_Te\mu _{_{PT}}}{\hbar ^2}}\;\left( 
{\frac{E_x}{\hbar c}}\right) ^\lambda \;\sqrt{\lambda +1}\;\exp \left\{
-i\mu \phi \right\} \;\Omega _\mu (q)  \nonumber \\
&\times &G_{\pi \lambda \mu }({\frac cv})\left\langle JM\left| {\cal M}_{\pi
\lambda ,-\mu }\right| J_0M_0\right\rangle \ ,
\end{eqnarray}
\begin{equation}
\Omega _\mu (q)=\int_0^\infty db\;b\;J_\mu (qb)K_\mu \left( {\frac{E_xb}{%
\gamma \hbar v}}\right) \exp {i\chi (b)}\ ,
\end{equation}
$J_\mu (K_\mu )$ is the cylindrical (modified) Bessel function of order $\mu 
$, and the functions $G_{\pi \lambda \mu }(c/v)$ are tabulated in ref. \cite
{AW79}.

The cross sections for Coulomb plus nuclear excitation are given by 
\begin{equation}
{\frac{d\sigma_{\lambda}^{J}}{\; d\Omega dE_x}} ={\frac{1}{2J_0+1}} \;
\sum_{M_0, \ M} \left| f_C^{\lambda\mu} + f_N^{\lambda\mu} \right|^2
\end{equation}

Further simplifications can be obtained by noticing that, since the M3Y
interaction does not depend on isospin, the nuclear isovector excitations
are absent from $f_N$. This is well justified, since even in the case when
isovector excitations can be reached, they are of minor importance compared
to the isoscalar ones \cite{Sat72}. Thus, the nuclear excitation of
isovector dipole modes ($\lambda=1$) will be suppressed and we can neglect
this multipolarity (also because intrinsic isoscalar dipole excitations do
not exist) in the sum of eq. (4.8). Moreover, angular momentum selection
rules imply that the $J=1^+$ and $J=3^+$ do not contribute to $\lambda=0$
excitation amplitudes. Also, non-resonant $p_{1/2}$ waves cannot be reached
via $\lambda=0$ excitation. Thus, nuclear monopole ($\lambda=0$) excitations
will also be absent from this sum.

For the Coulomb amplitude we will consider $E1$, $E2$, and $M1$ excitations
(monopole, $E0$, excitations cannot be achieved in Coulomb excitation). Thus
the cross section including both Coulomb and nuclear excitation becomes 
\begin{equation}
{\frac{d\sigma _{\pi \lambda }^J}{\;d\Omega dE_x}}={\frac 1{2J_0+1}}%
\;\sum_{M_0,\ M}\left[ \left| f_C^{(E1)}\right| ^2+\left| f_C^{(M1)}\right|
^2+\left| f_N^{(2)}+f_C^{(E2)}\right| ^2\right] \ .
\end{equation}
The nuclear-Coulomb interference will only appear for quadrupole excitations.

\section{Results}

In figure 3(a) we plot the cross section for the nuclear excitation of $^8B$
projectiles incident on $Pb$ targets at 50 MeV/nucleon as a function of the
scattering angle in degrees, and for a relative energy, $E_{rel}$, between
the proton and the $^7Be$ fragment equal to 0.2 MeV ($E_x=E_{rel}+0.14\ MeV$%
). The excitation cross sections to the p- and f-waves and to the $3^{+}$
resonance are displayed. Since the $1^{+}$ state has a very small width, it
does not contribute appreciably for this excitation energy. Figure 3(b)
shows the same calculation, but for a relative energy of 1.2 MeV. The basic
feature in these being an oscillatory pattern, characteristic of diffraction
by a strong absorptive object. The measured angular distributions on the
other hand are rather flat in the angular range of interest, suggesting
small nuclear contribution(s).

In figure 4 we plot the angular integrated nuclear excitation cross section
as a function of the relative energy and for the bombarding energy of 50
MeV/nucleon. We see that most of the excitation cross section goes to the $%
1^+$ and $3^+$ state; the excitation of the p- and f-waves being of much
smaller magnitude. We observe that the nuclear interaction broadens the
width of the $1^+$ and $3^+$ resonances (e.g., for the $1^+$ resonance, $%
\Gamma\approx 50\ keV$).

In figure 5(a) and we show the Coulomb excitation cross sections of $^8B$
projectiles incident on $Pb$ targets at 50 MeV/nucleon as a function of the
scattering angle in degrees, and for a relative energy, $E_{rel}$, between
the proton and the $^7Be$ fragment equal to 0.2 MeV ($E_x=E_{rel}+0.14\ MeV$%
). In figure 5(b) we use $E_{rel}$=1.2 MeV. We notice that the E1 excitation
dominates at low angles and that the E2 excitation becomes as strong as the
E1 mode at larger angles. The peak value occurs at $\theta \sim 0.4^{\circ }$
where the E1 cross section is more than one order of magnitude bigger that
the E2, and more than two orders of magnitude bigger than the M1 excitation.
Plots for 250 MeV/nucleon are presented in figs. 6(a) and 6(b),
respectively. It is also interesting to compare these calculations with the
semiclassical formula presented in ref. \cite{Ber95} for the E1 excitation
mode, which is much easier to use. As we see in figure 7, the cross section
is very well reproduced by the semiclassical formula. Diffraction effects
are of minor relevance and only introduce wiggles in the cross section
around the semiclassical results.

In figure 8(a) we show the Coulomb excitation cross section integrated over
angles for 50 MeV/nucleon, as a function of the relative energy of the
fragments. Figure 8(b) is for 250 MeV/nucleon. Note that E2 excitation are
dominant at large angles and hence the E2 mode contributes one fifth of the
total cross section even though it is negligible at small angles (see
below). And the M1 excitation is only relevant around the $1^+$ resonance, $%
E_{rel}\sim 600\ keV$. Comparing this figure with figure 4 we see that the
nuclear contribution to the total excitation cross section is about 3 orders
of magnitude smaller than the Coulomb excitation cross sections for the
whole spectrum. This is a very important result, since one can neglect the
nuclear excitation cross sections for practical purposes. The same applies
for 250 MeV/nucleon.

In figs. 9(a), 9(b) and 9(c) we compare the results of our model with the
angular distributions of $^8B$ breakup on lead targets at 50 MeV/nucleon
measured by Kikuchi {\em et al.} \cite{Ki97} at RIKEN. We have used the
acceptance (efficiency) matrix as well as angular and energy averaging
procedures as discussed in Ref. \cite{Ki97} and provided by the RIKEN
collaboration \cite{Ki97}. We plot our predictions for the contributions of
the E1, E2, and of the nuclear excitation cross sections. The solid line is
the sum of all three contributions, and we note the rather good agreement
with the experimental data with E1 excitation solely. Our model suggest a
very negligible E2 contribution as was the conclusion of Kikuchi {\em et al.}
\cite{Ki97}, where upper limits on the E2 contribution were extracted. These
upper limits are indeed consistent with a preliminary analysis of the
previous RIKEN data \cite{Mo94} that we published earlier \cite{Ga95a}.

\subsection{Conclusions}

We have constructed a simple model for $^8B$ as well as a simple optical
model for the elastic scattering of $^8B$ plus $^{208}Pb$, and used it to
predict nuclear excitations as well as E1, E2 and M1 excitations of $^8B$ at
50 and 250 MeV/u. We show that nuclear excitations as well as E2 and M1
excitations are negligible for the most part of the data taken at RIKEN and
at GSI.

\bigskip\bigskip
\noindent{\bf Acknowledgments}

\medskip

This work was supported in part by the FAPERJ/Brazil, by the 
MCT/FINEP/CNPQ(PRONEX)/Brazil under
contract No. 41.96.0886.00, and by the USDOE Grant No. DE-FG02-94ER40870.

\bigskip
{\bf Figure Caption}\\

{\bf Fig. 1} - (a) Transition densities for the bound state of $^8B$ to the $%
J=1^+$, and $J=3^+$, resonances at $E_x=0.63$ MeV (dashed line), and $%
E_x=2.17$ MeV (solid line), respectively. These transition densities have
been integrated over the width of the resonances. Also shown in this figure
(dotted line) is the transition density obtained by the Tassie model. (b)
Transition densities from the ground state of $^8B$ to the non-resonant p-
(solid curve) and f-waves (dashed curve).

{\bf Fig. 2} - (a) Real part of the transition potentials for the nuclear
excitation of $^8B$ at 50 MeV/nucleon from the ground state to the
resonances at $J=1^+$ (solid line) and $J=3^+$ (dashed line), respectively.
They are compared with the Copenhagen model (dotted line) for the transition
potential. (b) Real part of the transition potentials for the nuclear
excitation of $^8B$ at 50 MeV/nucleon from the ground state to the
non-resonant p- and f-waves.

{\bf Fig. 3} - Cross section for the nuclear excitation of $^8B$ projectiles
incident on $Pb$ targets at 50 MeV/nucleon as a function of the scattering
angle in degrees, and for a relative energy, $E_{rel}$, between the proton
and the $^7Be$ fragment equal to 0.2 MeV. The excitation cross sections to
the p- (solid curve) and f-waves (dashed curves) and to the $3^+$ resonance
(dotted curves) are displayed. (b) Same as in (a), but for $E_{rel}=1.2$ MeV.

{\bf Fig. 4} - Angular integrated nuclear excitation cross section of $^8B$
projectiles incident on lead targets at 50 MeV/nucleon. The cross sections
for the excitation of the $1^+$ (solid line), $3^+$ (dashed line)
resonances, and to the $p_{1/2}$ wave (dotted line) are shown.

{\bf Fig. 5} - (b) Coulomb excitation cross sections of $^8B$ projectiles
incident on $Pb$ targets at 50 MeV/nucleon as a function of the scattering
angle in degrees, and for a relative energy, $E_{rel}$, between the proton
and the $^7Be$ fragment equal to 0.2 MeV. The excitation cross sections due
to M1- (dashed-dotted), E2- (dotted curve), E1- (dashed curve) excitation modes are shown.
The solid curve is the sum of all these contributions. (b) Same as in (a),
but for $E_{rel}=1.2$ MeV.

{\bf Fig. 6} - Same as in figure 5, but for 250 MeV/nucleon.

{\bf Fig. 7} - Comparison between the quantal (solid curves) and the
semiclassical (dashed curves) calculations of Coulomb excitation of $^8B$
projectiles incident of lead targets for $E_{rel}=1.2$ MeV at 50 MeV/nucleon
and 250 MeV/nucleon, respectively.

{\bf Fig. 8} - (a) Coulomb excitation cross section of $^8B$ projectiles
incident of lead targets at 50 MeV/nucleon as a function of the relative
energy of the fragments, and for the M1- (dotted curve), E2- (dashed-dotted
curve) and E1- (dashed curve) excitation modes. The solid curve is the sum
of all contributions. (b) Same as in (a), but for 250 MeV/nucleon.

{\bf Fig. 9} - Angular distributions of $^8B$ breakup on lead targets at 50
MeV/nucleon. Data are from Kikuchi {\em et al.} \cite{Ki97}. The separate
contributions of the E1, E2, and nuclear interaction are shown.

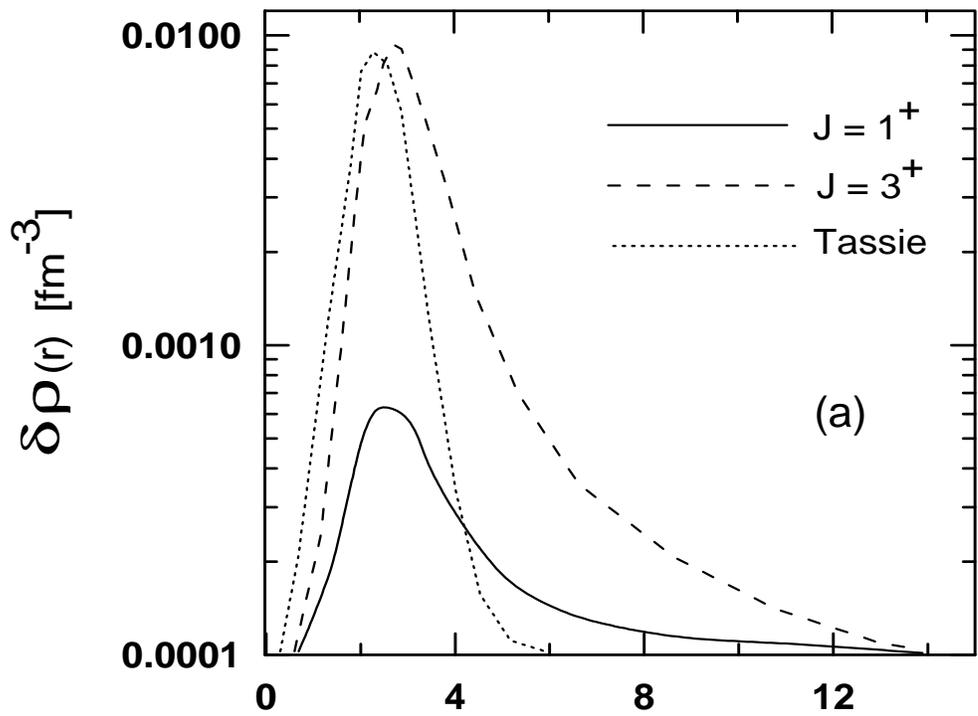
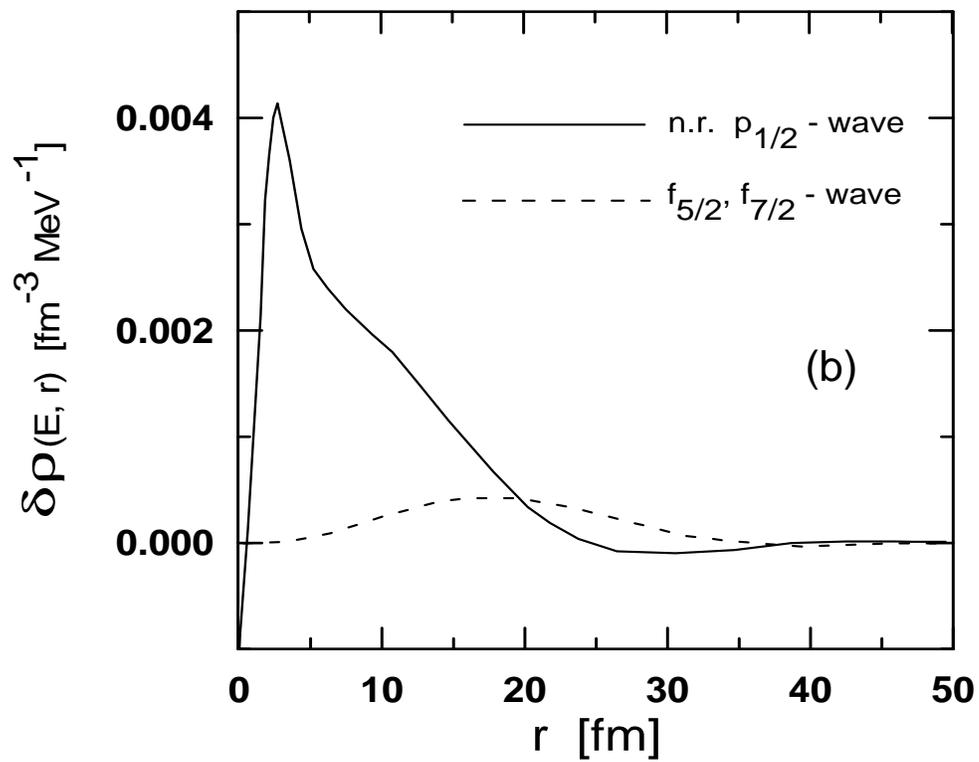
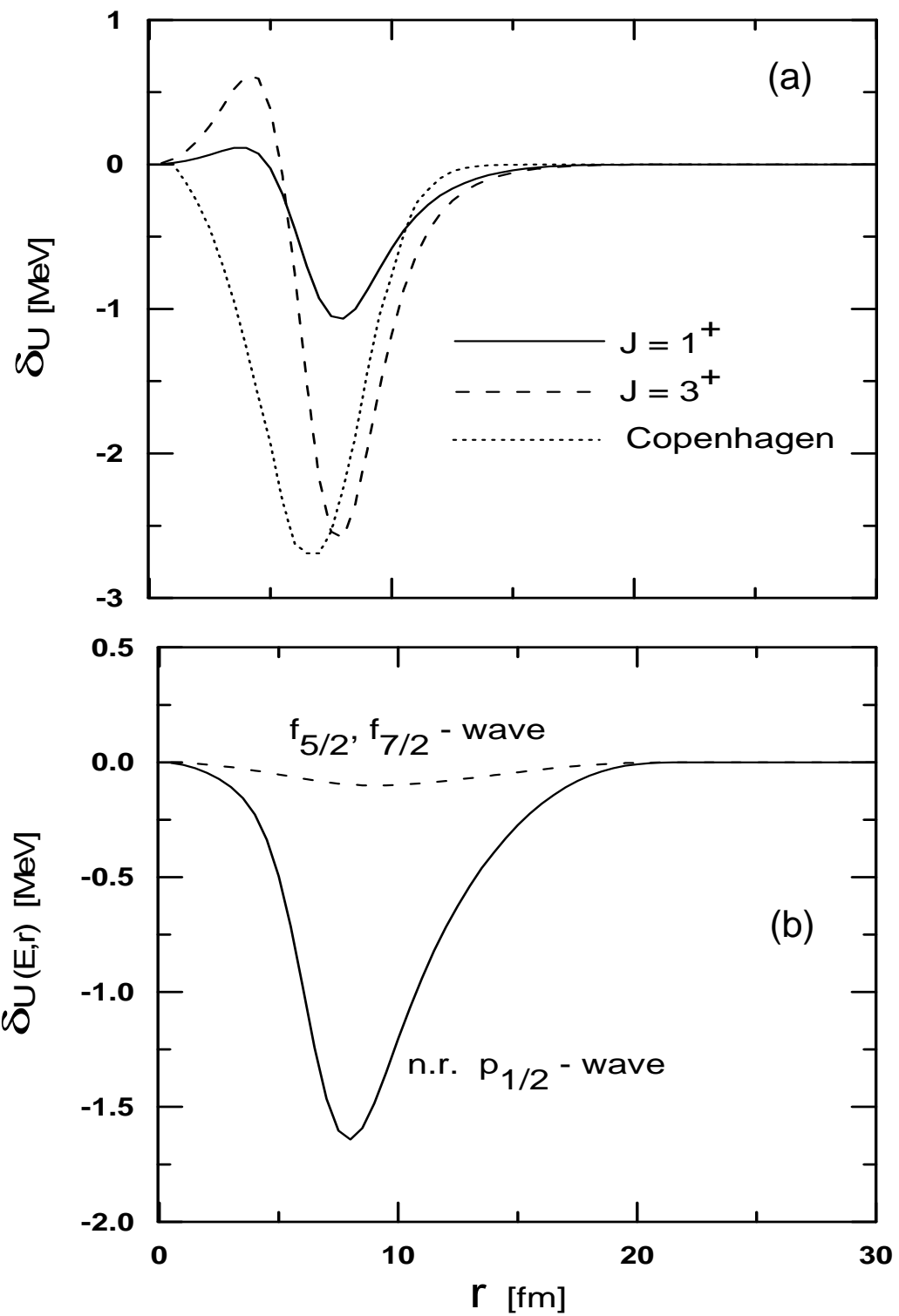
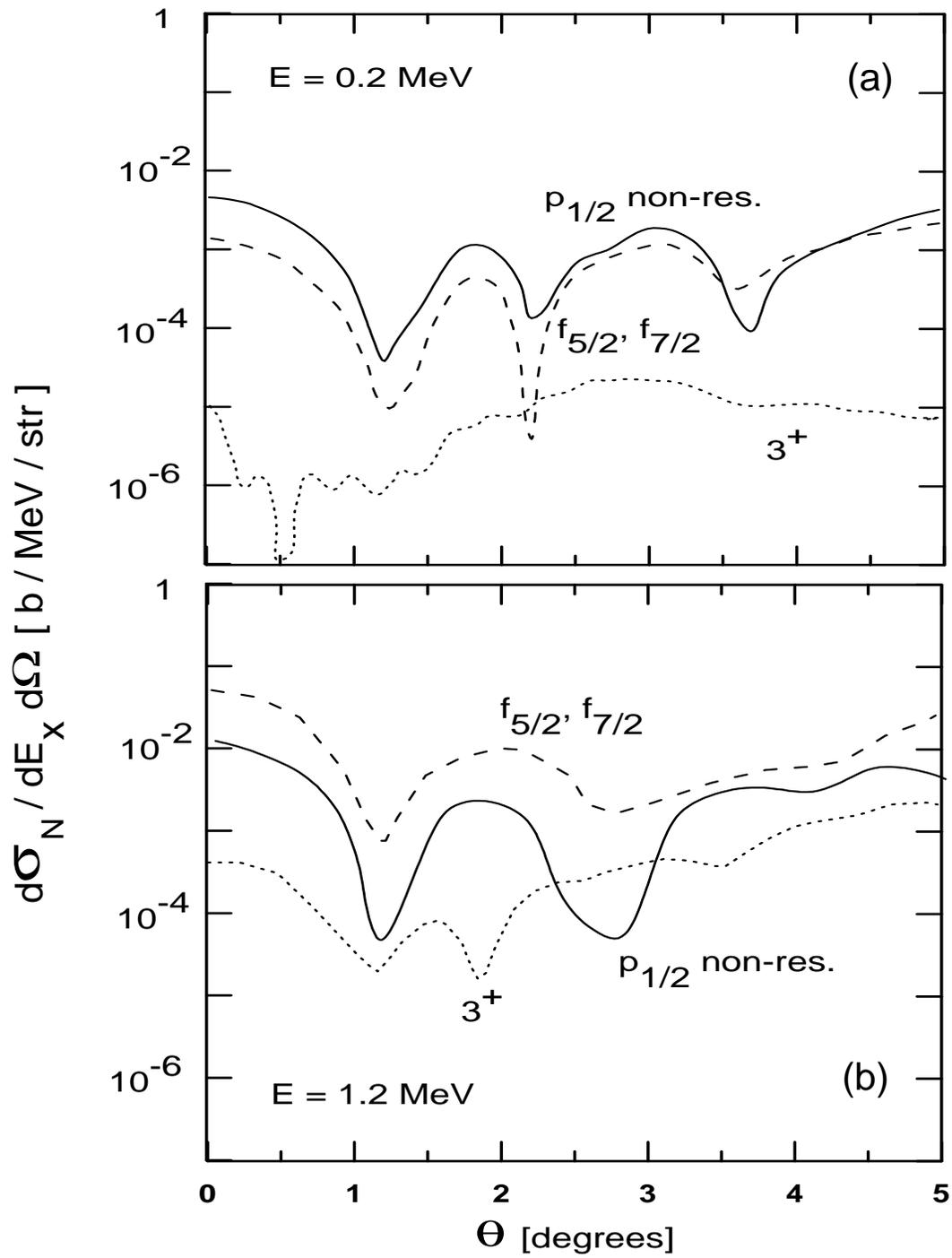
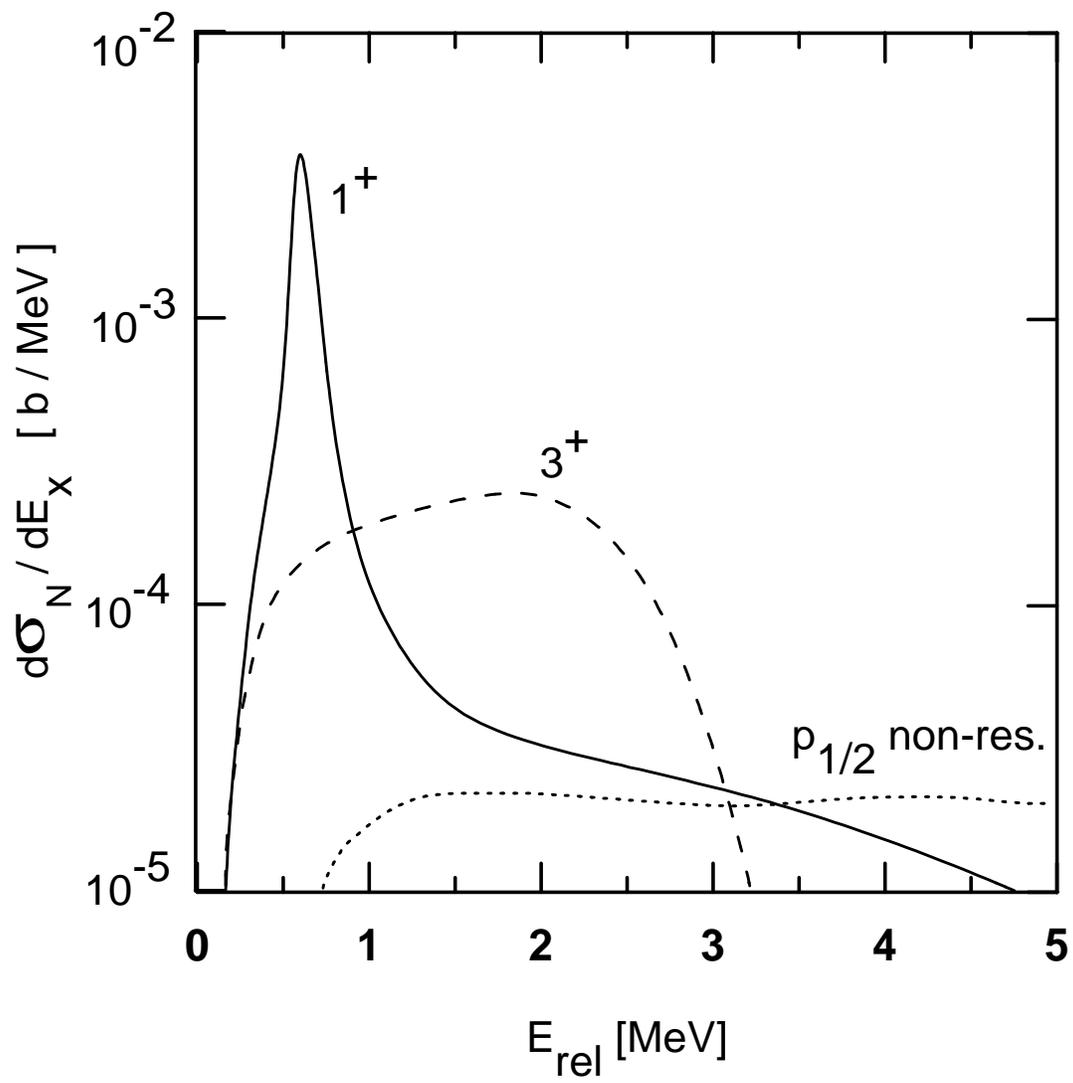
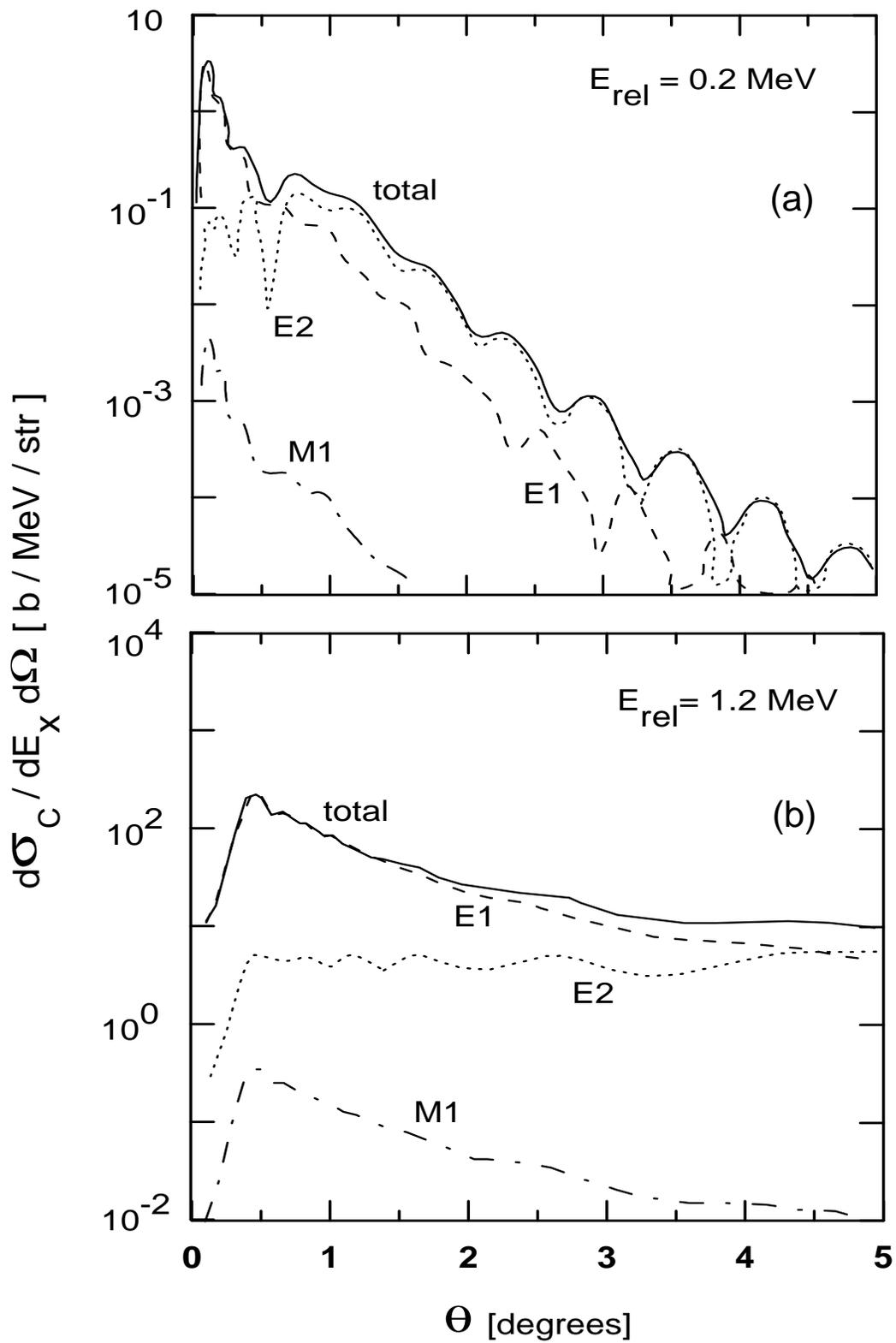
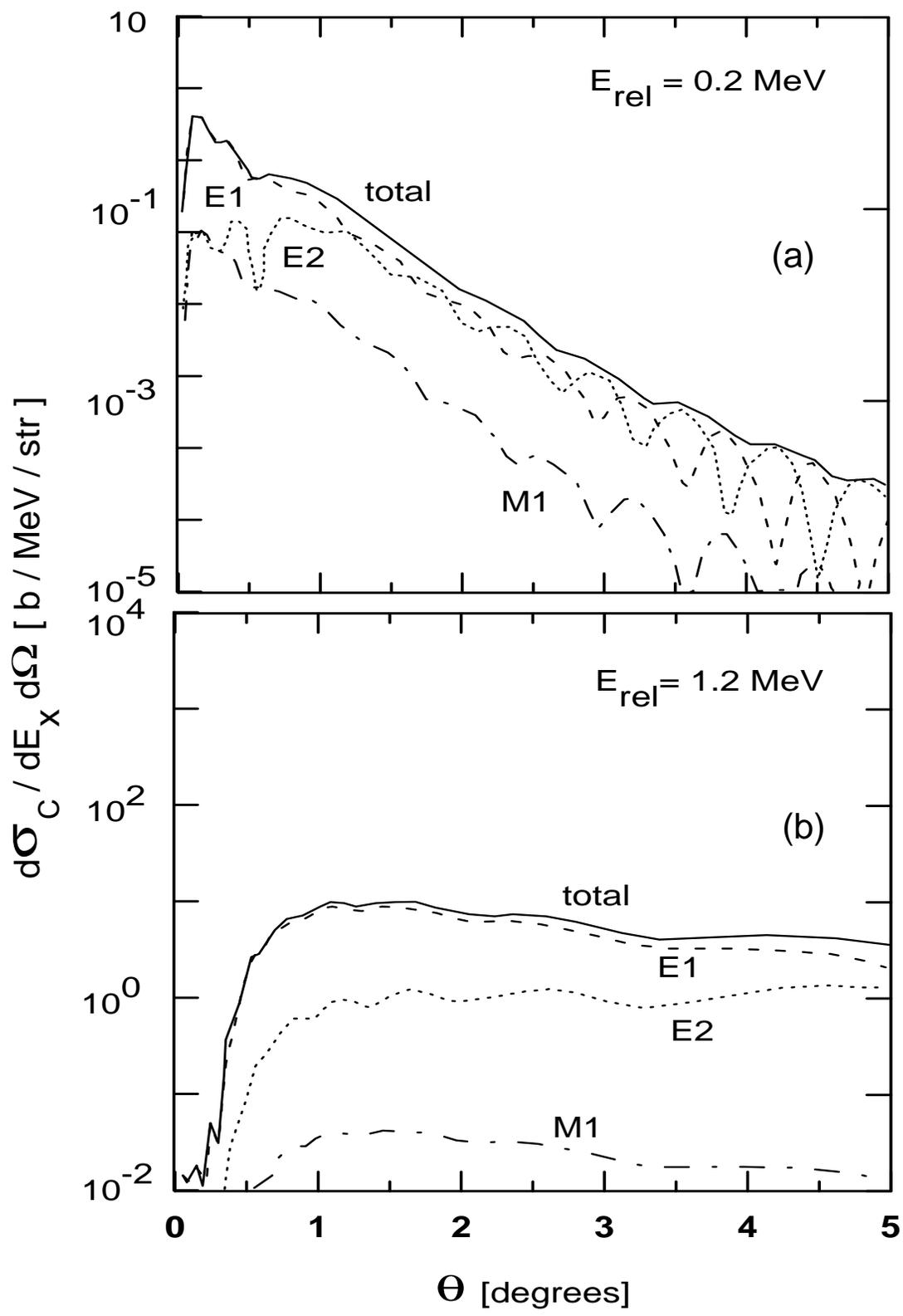
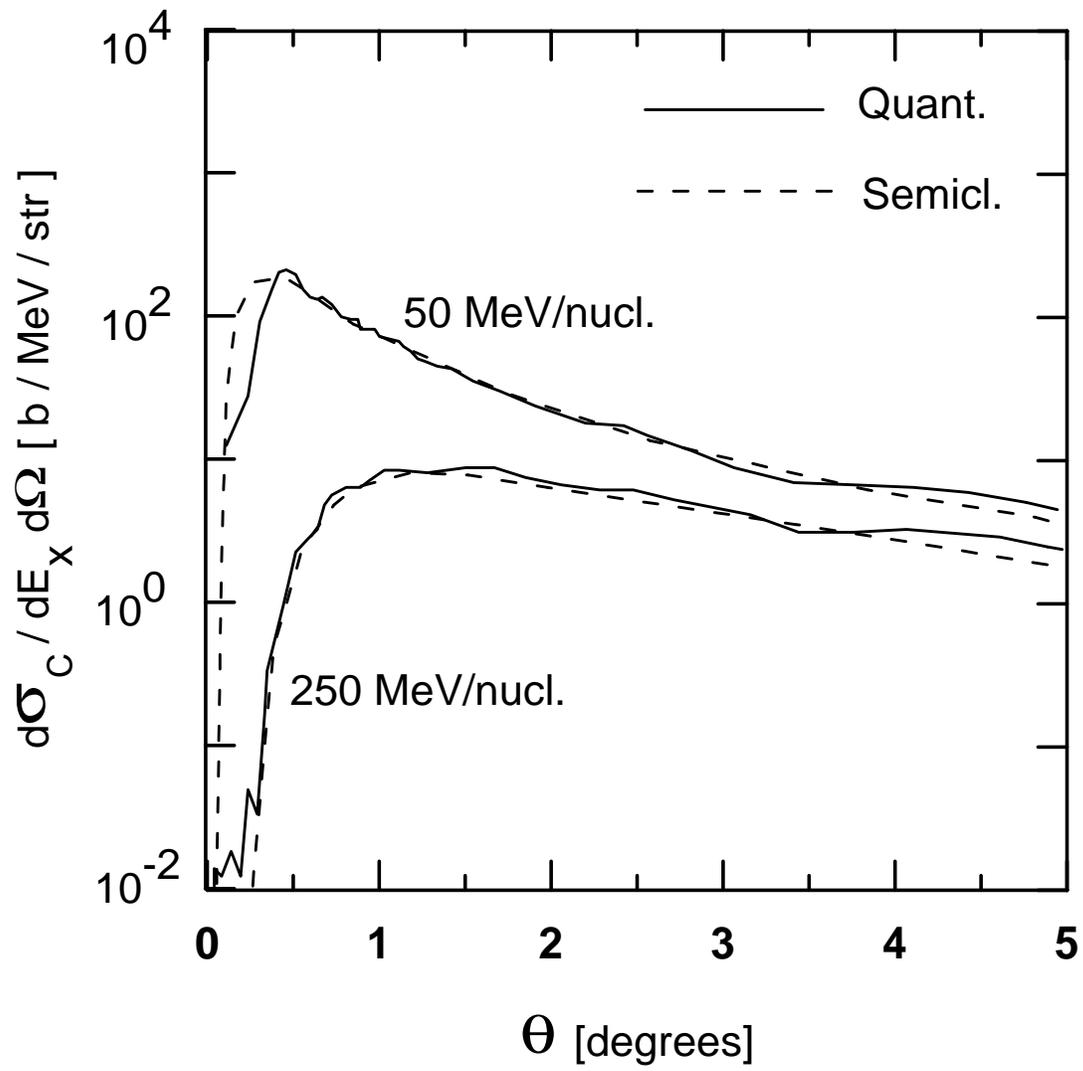
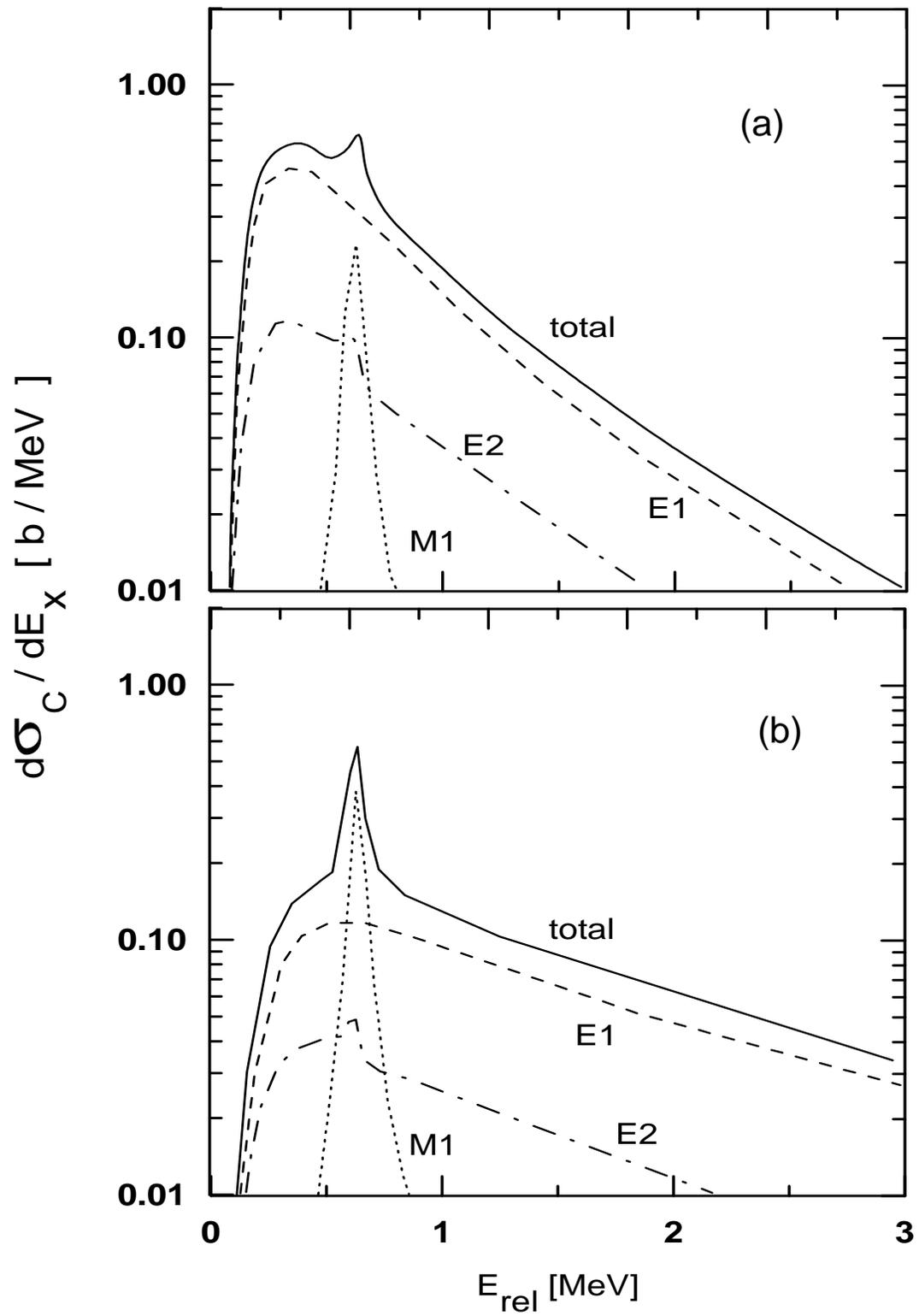
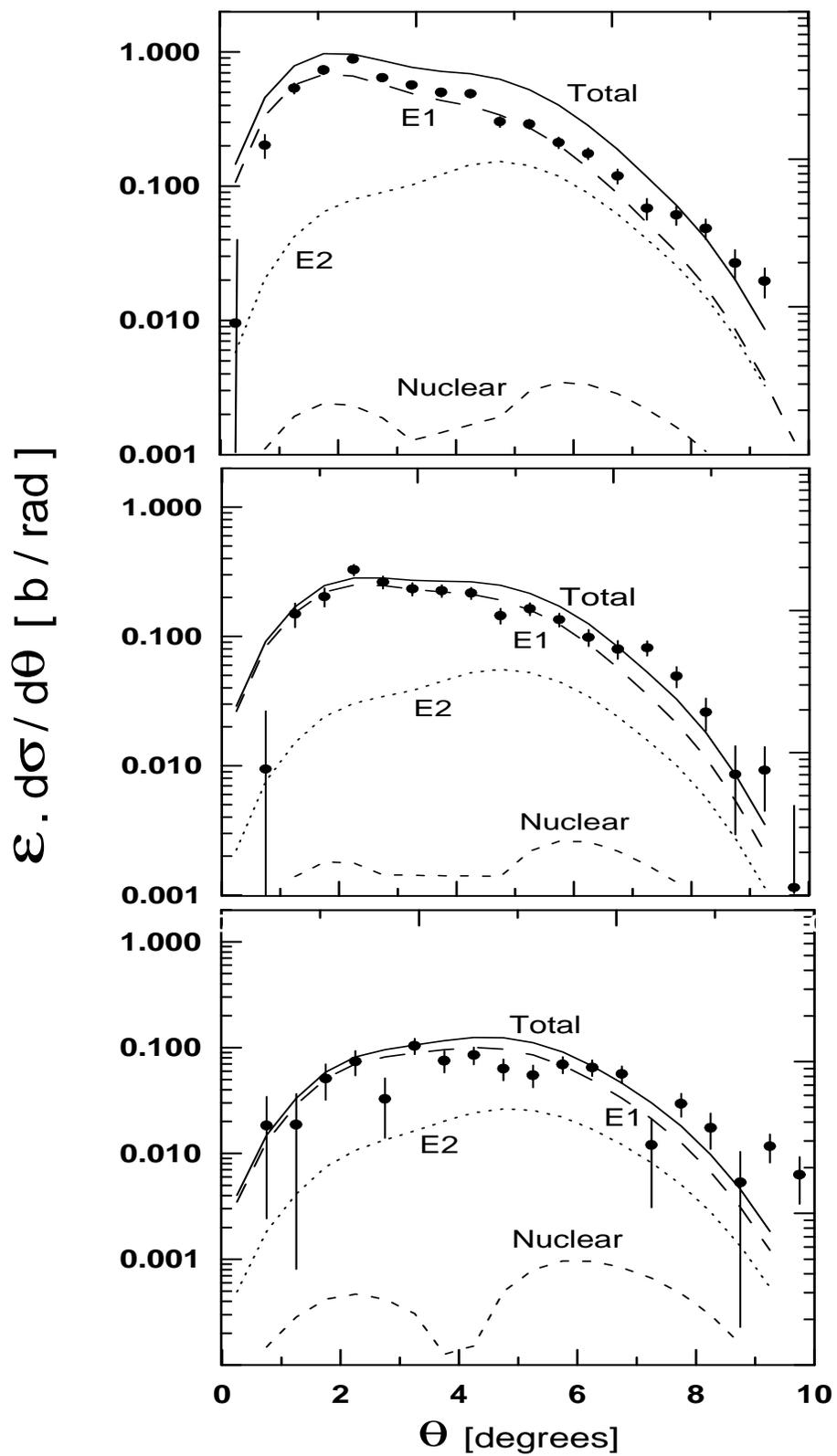
\end{document}